\documentclass[]{aastex63}
\usepackage{amsmath}

\submitjournal{ApJ}

\shorttitle{Magnetic switchback formation}
\shortauthors{Magyar et al.}

\begin{document}

\title{Could switchbacks originate in the lower solar atmosphere? \\ II. Propagation of switchbacks in the solar corona}

\correspondingauthor{Norbert Magyar}
\email{norbert.magyar@warwick.ac.uk}

\author[0000-0001-5731-8173]{Norbert Magyar}
\affiliation{Centre for Fusion, Space and Astrophysics, \\Physics Department, University of Warwick,\\ Coventry CV4 7AL, UK}
\affiliation{Centre for mathematical Plasma Astrophysics (CmPA),\\
KU Leuven, \\
Celestijnenlaan 200B bus 2400, B-3001 Leuven, Belgium}

\author[0000-0002-0061-5916]{Dominik Utz}
\affiliation{Computational Neurosciences, Neuromed Campus, Kepler University Hospital,\\ 4020 Linz, Austria}
\affiliation{Instituto de Astrofísica de Andalucía IAA-CSIC,\\
Glorieta de la Astronomía s/n,\\
18008 Granada, Spain}


\author[0000-0003-3439-4127]{Robertus Erd\'elyi}
\affiliation{Solar Physics \& Space Plasma Research Center (SP2RC),\\
School of Mathematics and Statistics, University of Sheffield, \\
Hicks Building, Hounsfield Road, S3 7RH, UK}
\affiliation{Dep. of Astronomy, E\"otv\"os Lor\'and Univ.,\\
P\'azm\'any P. s\'et\'any 1/A,\\
Budapest, H-1117, Hungary}
\affiliation{Gyula Bay Zolt\'an Solar Observatory (GSO),\\
Hungarian Solar Physics Foundation (HSPF), \\
Pet\H{o}fi t\'er 3., Gyula, H-5700, Hungary}

\author[0000-0001-6423-8286]{Valery M. Nakariakov}
\affiliation{Centre for Fusion, Space and Astrophysics, \\Physics Department, University of Warwick,\\ Coventry CV4 7AL, UK}
\affiliation{School of Space Research, Kyung Hee University, Yongin, 17104, \\Republic of Korea}

\begin{abstract}
The magnetic switchbacks observed recently by the Parker Solar Probe have raised the question about their nature and origin. One of the competing theories of their origin is the interchange reconnection in the solar corona. In this scenario, switchbacks are generated at the reconnection site between open and closed magnetic fields, and are either advected by an upflow or propagate as waves into the solar wind. \par 
In this paper we test the wave hypothesis, numerically modelling the propagation of a switchback, modeled as an embedded Alfv\'en wave packet of constant magnetic field magnitude, through the gravitationally stratified solar corona with different degrees of background magnetic field expansion. While switchbacks propagating in a uniform medium with no gravity are relatively stable, as reported previously, we find that gravitational stratification together with the expansion of the magnetic field act in multiple ways to deform the switchbacks. These include WKB effects, which depend on the degree of magnetic field expansion, and also finite-amplitude effects, such as the symmetry breaking between nonlinear advection and the Lorentz force. In a straight or radially expanding magnetic field the propagating switchbacks unfold into waves that cause minimal magnetic field deflections, while a super-radially expanding magnetic field aids in maintaining strong deflections. Other important effects are the mass uplift the propagating switchbacks induce and the reconnection and drainage of plasmoids contained within the switchbacks. In the Appendix, we examine a series of setups with different switchback configurations and parameters, which broaden the scope of our study.
\end{abstract}


\section{Introduction} \label{sec:intro}

Among the recent breakthrough findings of the Parker Solar Probe (PSP), the abundance of strong localized folds or kinks in the magnetic field, i.e. the so-called switchbacks or spikes, is in the spotlight \citep{2019Natur.576..228K,2020ApJS..246...45H}. Switchbacks come in a wide range of deflection angles, from a few degrees to a full reversal. The distribution of detected deflection angles resembles a power-law without specific populations present \citep{2020ApJS..246...39D}. There is preliminary evidence that switchbacks become more common with increasing distance from the Sun \citep{2021MNRAS.501.5379M}. Nevertheless, the formation mechanism(s) of switchbacks, and whether they originate in the lower solar atmosphere or locally in the solar wind are intensively debated. Currently, the most popular lower solar atmospheric formation mechanism is interchange reconnection in the solar corona \citep{2004JGRA..109.3104Y,2005ApJ...626..563F,2020ApJ...894L...4F}, in which switchbacks form at the reconnection sites between the open and closed magnetic fluxes. Some studies suggest that reconnection results in magnetic flux ropes ejected by the outflow that is generated \citep{2020arXiv200905645D}, while others suggest that reconnection generates propagating wave packages of either Alfv\'enic  \citep{2020arXiv200909254H} or fast magnetosonic \citep{2020ApJ...903....1Z} nature. On the other hand, proponents of the in-situ solar wind origin of switchbacks argue that the ensuing turbulent dynamics can readily generate these structures \citep{2020ApJ...891L...2S,2021arXiv210109529S}. A recent phenomenological model claims that switchbacks simply evolve as a result of velocity shear in the solar wind \citep{2021arXiv210203696S}. Therefore, the question about the origin and nature of switchbacks is still not concluded. Here the term `Alfv\'enic' refers to the highly correlated magnetic field and velocity perturbations (see Eq.~\ref{correl}), a property of Alfv\'en waves often measured in the solar wind, and to the unperturbed total magnetic field magnitude \citep[e.g.,][]{2020ApJ...904..195Y}. In \citet{2021ApJ..M} (in the following Paper I), we have examined whether flows in and around the footpoints of intense magnetic flux elements in the photosphere, e.g. associated with magnetic bright points, can generate switchbacks. This Paper I study was motivated by the conclusion of \citet{2020ApJS..246...39D} that switchbacks most likely originate deep in the corona, and additionally, to test an alternative, photospheric scenario for their origin. While we demonstrate in Paper I that flows, both upflows and shear flows, readily generate highly deflected magnetic fields and even full reversals in the chromosphere, these strongly folded structures appear to be confined to the lower solar atmosphere and are unable to enter the corona. \par 
Following our previous paper, we now continue the investigation of whether switchbacks could originate in the lower solar atmosphere, specifically whether Alfv\'enic wave packages originating in the lower solar corona are able to propagate out to distances that were measured by PSP. Previous studies employing numerical simulations have demonstrated the generation of propagating Alfv\'enic spikes \citep{2020arXiv200909254H}, or flux ropes ejected upwards \citep{2020arXiv200905645D} through interchange reconnection, resembling switchbacks. However, the subsequent evolution and propagation of the generated switchbacks was not followed, therefore it remains unclear whether these features are stable for long enough to propagate out into the solar wind.
\citet{2018ApJ...852...98W} points out that the highly twisted flux ropes resulting from interchange reconnection are quickly relaxed towards a potential un-kinked field. In order to investigate the stability of propagating Alfv\'enic switchbacks, \citet{2020ApJS..246...32T} ran a simulation of an embedded switchback propagating along a straight magnetic field, in a periodic domain without gravity. Their conclusion was that switchbacks originating in the corona could be stable for long enough to travel out to PSP-scanned distances, provided that the background solar wind does not possess itself significant inhomogeneities. This finding was a big leap forward for establishing switchback stability compared to the previous study of e.g. \citet{2006GeoRL..3314101L}, on which the work of \citet{2020ApJS..246...32T} is based. The difference between the two studies is that \citet{2006GeoRL..3314101L} employed a non-constant magnetic field switchback, that is, one with total pressure imbalance, which lead to the rapid untangling of the switchback. It has to be specified that strong folds in the magnetic field that present magnetic pressure gradients are necessarily unstable, as gas pressure cannot provide a constant total pressure configuration in the low plasma-beta corona. Therefore, constant magnetic field magnitude swithcbacks are  possibly the only structures that might remain stable for any duration of time in the corona. \par 
In this paper, we further develop the models of \citet{2006GeoRL..3314101L} and \citet{2020ApJS..246...32T}, by embedding an Alfv\'enic switchback in the gravitationally stratified lower corona with varying degrees of magnetic field expansion and follow its evolution as it propagates upwards. Additionally, in the Appendix we present a series of simulations that strengthen our conclusions by exploring a range of representative setups. In general, we find that stratification and expansion have a strong impact on the stability of the propagating switchback, with strong deformations observed.
The paper is organized as follows. In the following Section~\ref{sec:model} we describe the numerical model. In Section~\ref{sec:result}, we present the results for the main simulation setup, along with some discussion of the results. In Section~\ref{sec:concl}, we conclude the results and comment on possible future work and caveats of the present study. Finally, in the Appendix, we present the results of various related additional simulations, e.g., runs without gravity, with different switchback shapes, and switchback embedded in the chromosphere. 

\section{Numerical model} \label{sec:model}

\begin{figure}[h]
    \centering     
        \begin{tabular}{@{}cc@{}}
        \includegraphics[width=0.4\textwidth]{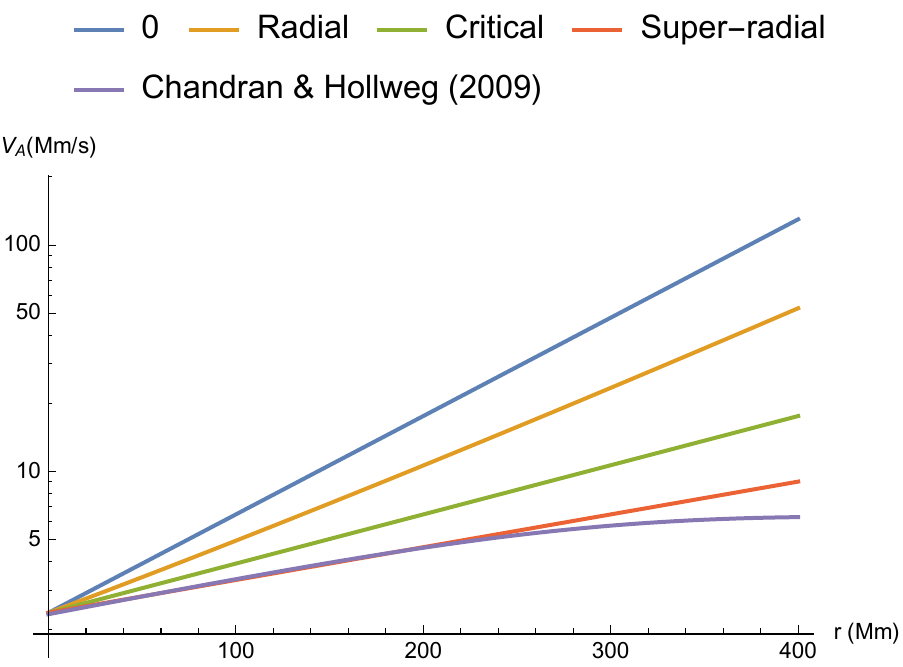}  
        \includegraphics[width=0.4\textwidth]{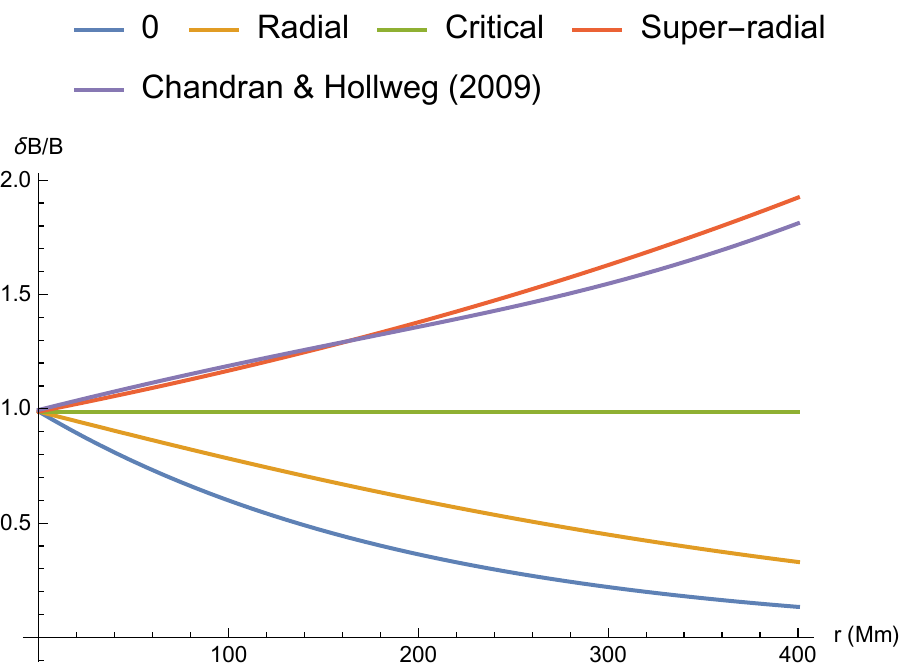} \\
       \end{tabular}   
        \caption{\textit{Left}: Alfv\'en speed as a function of radial distance. \textit{Right}: Normalized relative magnetic field perturbation amplitude, according to WKB theory \citep[e.g.,][]{2001A&A...374L...9M}, as a function of radial distance. Values are shown for the four different expansions considered, "0" ($l_B = \infty$), "Radial" ($l_B = 400\ \mathrm{Mm}$), "Critical" ($l_B = 200\ \mathrm{Mm}$), and "Super-radial" ($l_B = 150\ \mathrm{Mm}$) compared to the profile in \citet{2009ApJ...707.1659C}.}
        \label{fig1b}
\end{figure}

\begin{figure}[t]
    \centering     
     \begin{tabular}{@{}cc@{}}
        \includegraphics[width=0.4\textwidth]{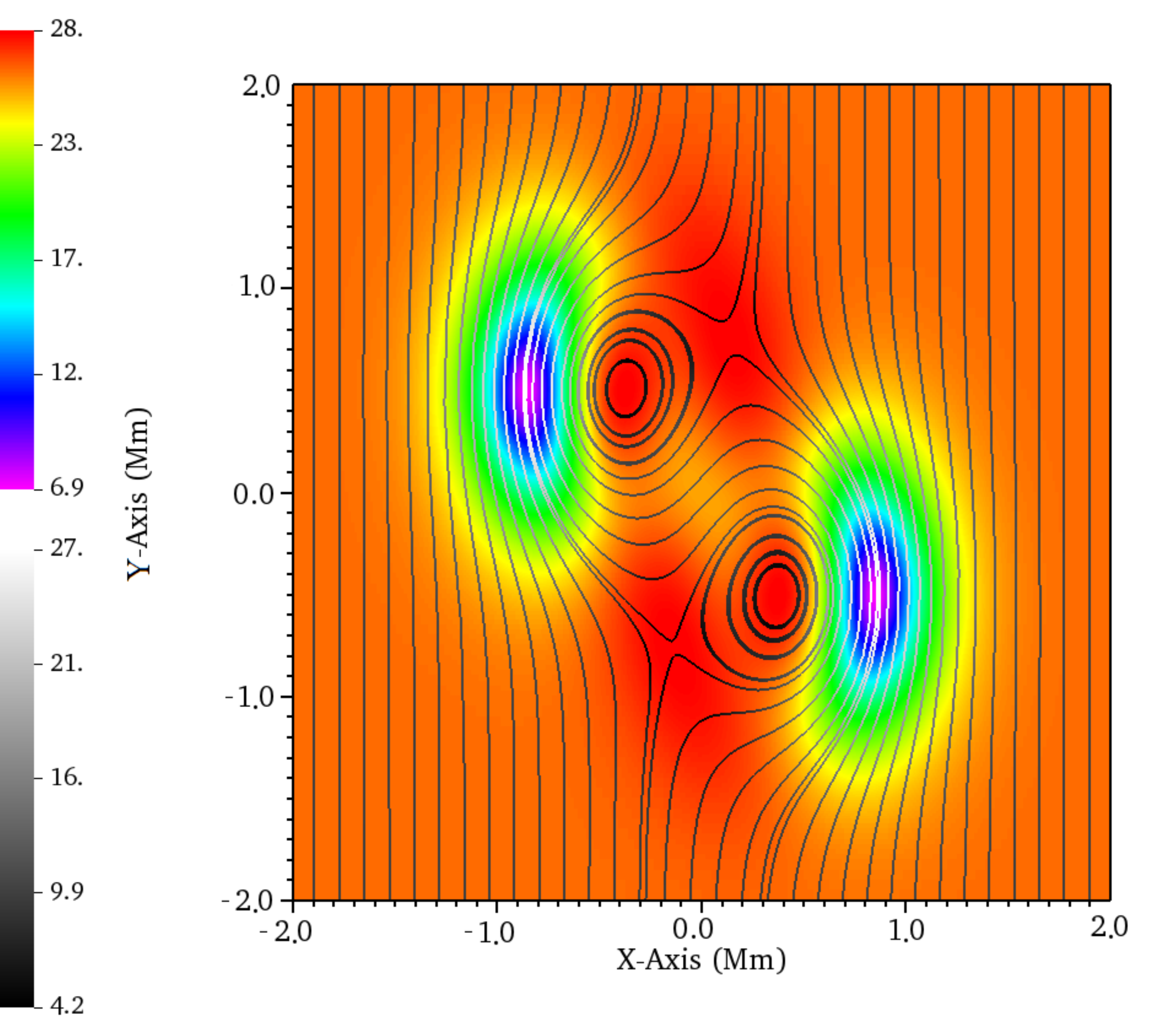}  
        \includegraphics[width=0.42\textwidth]{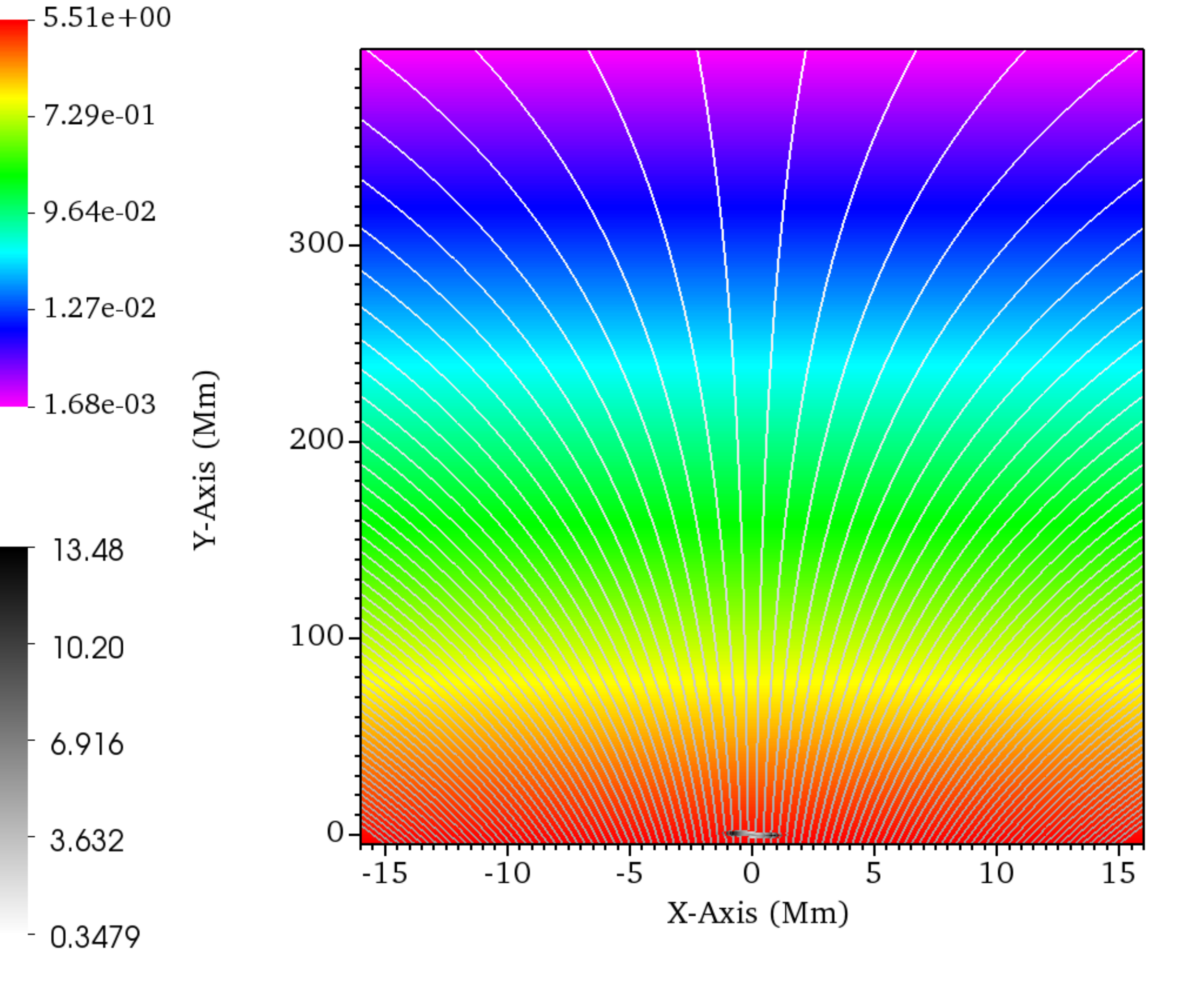} \\
       \end{tabular}    
        \caption{\textit{Left}: Snapshot of the initial configuration of the magnetic field with no expansion, centered on the embedded switchback. Field lines of the in-plane ($x-y$-plane) magnetic field are shown, with the gray-scale bar representing the total magnetic field magnitude. The color plot shows the out-of-plane component of the magnetic field. Any deviation from symmetry is a visualization artifact. \textit{Right}: Snapshot of the whole domain for the super-radially expanding setup ($l_B = 150\ \mathrm{Mm}$). The color plot shows the background density logarithmically, in units of $2.34 \cdot 10^{-12}\ \mathrm{kg\ m^{-3}}$. Field lines of the in-plane ($x-y$-plane) magnetic field are shown, with the gray-scale bar representing the in-plane magnetic field magnitude. Magnetic field is in units of Gauss.}
        \label{fig1}
\end{figure}

We run 2.5D (two spatial coordinates and
three-dimensional fields) MHD numerical simulations with Cartesian geometry using \texttt{MPI-AMRVAC} \citep{2014ApJS..214....4P,2018ApJS..234...30X}. A finite-volume three-step \texttt{hll} solver is employed, with \texttt{woodward} slope limiter. The solenoidality of the magnetic field is maintained by using a constrained transport method. The initial condition consists of a gravitationally-stratified corona in equilibrium, with the $y$-axis directed vertically, in the direction opposite to the gravity. There are no background flows. The equilibrium density is uniform in the horizontal direction. Therefore, the density and pressure are given by:
\begin{equation}
 \begin{array}{lr}
    \rho(y) = \rho_0\ \exp\left(-\frac{g}{R_{sp}T}\left(\frac{R_\odot^2}{R_\odot+y}-R_\odot\right)\right),\\
     p(y) = \rho(y) R_{sp} T,
 \end{array}
 \label{initcond}
\end{equation}
where $\rho_0 = 10^{-11}\ \mathrm{kg/m^3}$ is the density at $y=0$, which is $5\ \mathrm{Mm}$ higher than the base of the simulation domain, $R_\odot$ is the radius of the Sun, $T = 1\ \mathrm{MK}$ is the uniform temperature in the corona, $R_{sp} = k_{B}/m$ is the specific gas constant, where $k_B$ is Boltzmann's constant and $m = 0.62\ m_p$ is the mean mass per particle for coronal abundances, with $m_p$ the proton mass. The background magnetic field is given by
\begin{equation}
   \begin{array}{ll}
      B_{0x}(x,y) =& B_0 \sin\left(\frac{x}{l_B} \right) \exp\left(-\frac{y}{l_B}\right), \\
      B_{0y}(x,y) =& B_0 \cos\left(\frac{x}{l_B} \right) \exp\left(-\frac{y}{l_B}\right), \\
      B_{0z} =& \sqrt{B_t^2 - B_0^2},
   \end{array}
\end{equation} 
where $B_0 = 10\ G$, $B_t = 28\ G$, and $l_B$ is the  magnetic scale height, that is, the parameter controlling the expansion of the magnetic field. By varying the value of $l_B$, we study the effect of expansion. The magnetic field expansion is varied to account for multiple expansion regimes of the magnetic field: no expansion, radial expansion $(1/r^2)$, critical expansion, and super-radial expansion.
Despite the varying degrees of expansion, a common property across the setups is an increasing Alfv\'en speed with radial distance. Such background coronal solutions are commonly employed in models with expanding magnetic fields, including super-radial expansion \citep[e.g.,][]{2009ApJ...707.1659C,2019JPlPh..85d9009C}, the Alfv\'en speed attaining a maximal value around $1.5-2\ \mathrm{R_\odot}$. See Fig.~\ref{fig1b} for the Alfv\'en characteristics of the different expansion profiles.
The difference between the super-radial expansion in our setup and the profile in \citet{2009ApJ...707.1659C} originates in the background flow accounted for in the latter, which we neglect. Including an equilibrium with background flow would require an external heating function.
An Alfv\'enic switchback is embedded in the lower part of the simulation domain, centered around $(x,y) = (0,0)$, of the form given in \citet{2020ApJS..246...32T}. The components of the magnetic field perturbation are given by
\begin{equation}
    \begin{array}{ll}
         B_x(x,y) =& A \left(\frac{2 (y-y_2) \exp\left(-\frac{(x-x_2)^2}{l_x^2}-\frac{(y-y_2)^2}{l_y^2}\right)}{l_y^2}-\frac{2 (y-y_1)
   \exp\left(-\frac{(x-x_1)^2}{l_x^2}-\frac{(y-y_1)^2}{l_y^2}\right)}{l_y^2}\right),  \\
         B_y(x,y) =& -A \left(\frac{2 (x-x_2) \exp\left(-\frac{(x-x_2)^2}{l_x^2}-\frac{(y-y_2)^2}{l_y^2}\right)}{l_x^2}-\frac{2 (x-x_1)
   \exp\left(-\frac{(x-x_1)^2}{l_x^2}-\frac{(y-y_1)^2}{l_y^2}\right)}{l_x^2}\right), \\ 
         B_z(x,y) =& B_{0z} - \sqrt{B_{t}^2-B_x(x,y)^2-B_y(x,y)^2}.
    \end{array}
    \label{bpert}
\end{equation}
Here, $A$ is a coefficient that determines the amplitude of the perturbation. The embedded switchback is of constant magnetic pressure. The position, size and shape of the switchback are determined by setting $x_1 = 0.5, x_2 = -0.5, y_1 = -0.5, y_2 = 0.5, l_x = 0.5, l_y = 0.75\ (\mathrm{Mm})$. See Fig.~\ref{fig1} for depictions of the initial magnetic field perturbation for $A = 10\ \mathrm{G\cdot Mm}$.
Note that for this amplitude, the switchback solution is presenting regions of closed magnetic field, which we refer to as plasmoids. 
The initial velocity perturbation of the embedded switchback is obeying upward-propagating Alfv\'enic correlation, i.e. it renders the switchback an upward-propagating Alfv\'enic wave-packet:
\begin{equation}
    \delta\mathbf{v} = - \delta\mathbf{B} \left(\mu \rho(y) \right)^{-1/2},
\label{correl}
\end{equation}
where $\delta\mathbf{B}$ is the magnetic field perturbation vector with components given above in Eq.~\ref{bpert}. Alfv\'en wave-packets of magnetic field with a constant magnitude are referred to as spherically-polarized \citep[e.g.,][]{1998JGR...103..335V,2012PhRvL.109w1102R}. Plasma beta, the ratio of gas to magnetic pressure, is decreasing vertically from a bottom value of $\beta \approx 0.06$. The extents of the numerical domain are $400\ \mathrm{Mm}$ and $10 + 3 \cdot 10^3/l_B\ \mathrm{Mm}$ in the vertical and horizontal directions, respectively. Extending the domain horizontally for expanding magnetic fields is required as otherwise the switchback would expand beyond the horizontal extent of the box. The base numerical grid consists of $1024 \times 64$ cells for no expansion ($l_B = \infty$). For simulations with non-zero expansion the domain is extended in the horizontal direction while preserving the base resolution. We use seven additional levels of refinement, bringing the effective resolution to $65536 \times 8192$ cells (for $l_B = \infty$). The refinement criterion is a deviation of more than 5\% of the magnetic field components from their background values, limiting the maximal refinement to the extent of the switchback. At the top and bottom boundaries, the gravitational stratification of density and pressure are extrapolated from the values in the boundary cell. Otherwise, at all boundaries and for all variables, zero-divergence `open' conditions are set, allowing perturbations to freely leave the domain, if any.

\section{Results and Discussion} \label{sec:result}

\begin{figure}[h]
    \centering     
        \includegraphics[width=1.0\textwidth]{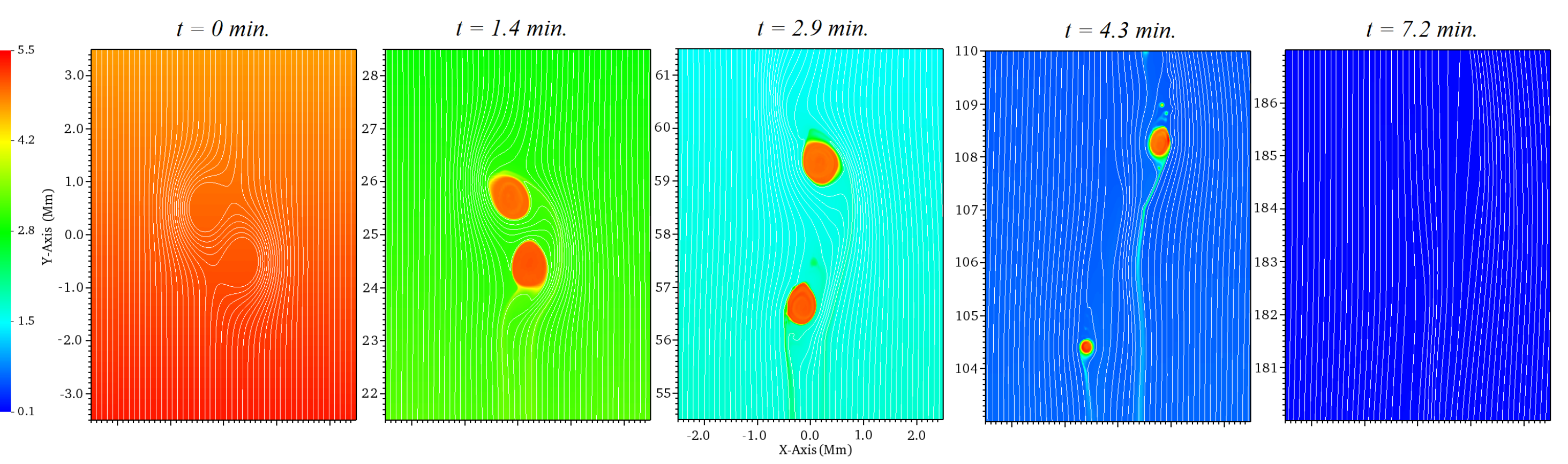} 
        \caption{A sequence of snapshots, showing the evolution of the switchback as it propagates, from left to right, is shown. The magnetic field is non-expanding ($l_B = \infty$). The color plots display density, in units of $2.34 \cdot 10^{-12}\ \mathrm{kg\ m^{-3}}$. Over-plotted are magnetic field lines (a zoom in for a clearer visualization). The leftmost plot depicts the initial condition, and following plots are at equal steps of $\approx 86\ \mathrm{s}$. }
        \label{fig2}
\end{figure}

The simulations are run until the propagating switchback reaches the upper vertical boundary at 400 Mm from the base, which renders the total simulation time dependent on the expansion of the magnetic field through its impact on the integrated Alfv\'en crossing time in the vertical direction. The main results are presented for a switchback of amplitude $A = 10\ \mathrm{G\cdot Mm}$, for different degrees of magnetic field expansion. In the Appendix, we briefly present results with switchbacks of small enough amplitude to not display plasmoids, among other setups. Compared to the relatively stable propagation of switchbacks in a homogeneous medium, as in \citet{2020ApJS..246...32T}, gravitational stratification and magnetic field expansion have a strong impact on the evolution of the switchbacks. One can distinguish between so-called WKB effects, which affect large-amplitude Alfv\'en waves essentially in the same way as they do small-amplitude linearized Alfv\'en waves \citep{1974JGR....79.2302B,1974JGR....79.1539H}, and effects that are only present due to the finite amplitude of the switchback. The relevant WKB effects that impact the evolution of the propagating switchback are presented first. As shown previously, across all setups with different degrees of expansion, the background Alfv\'en speed is increasing with height. Therefore, the leading part of the switchback is propagating slightly faster than the trailing part, resulting in the vertical stretching of the switchback. At the same time, as a consequence of wave energy flux conservation, the magnetic field perturbation is expected to decrease as $B \sim \rho^{1/4}$, per WKB approximation \citep[e.g.,][]{2000A&A...353..741N,2001A&A...374L...9M}. However, the important quantity here is the magnetic field perturbation amplitude relative to the background magnetic field, which varies considerably between the setups with different expansion rates. As one can see in Fig.~\ref{fig1b}, for expansion factors less than a critical expansion rate, including radial expansion, the switchback is expected to become less kinked, while for super-radial expansions greater than the critical expansion the switchback is expected to grow in amplitude. In this sense, the setup with critical expansion allows us to study the evolution of the switchback without the WKB effect due to wave energy flux conservation distorting its shape. First, we investigate the setups with no or radial expansion of the magnetic field. 
The evolution of the switchback propagating in a non-expanding magnetic field is depicted in Fig.~\ref{fig2}.
The switchback is untangled as it propagates, and results in perturbations with minimal deflections of the magnetic field. 
The evolution of the propagating switchback in the radially expanding magnetic field is similar, except that the dynamics observed at around $y = 100\ \mathrm{Mm}$ in Fig.~\ref{fig2} for the non-expanding setup are only observed around $y = 150\ \mathrm{Mm}$ with radial expansion. Thus, expansion of the magnetic field delays the untangling of propagating switchbacks, allowing them to retain their shape over longer distances. However, radial expansion is insufficient  in preventing their ultimate untangling. The added stability observed in the radially expanding setup is likely attributable to a shallower decrease in the relative perturbation amplitude with radial distance compared to the setup with no expansion, the WKB effect described above (see Fig.~\ref{fig1b}). \par
\begin{figure}[h]
    \centering     
        \includegraphics[width=0.65\textwidth]{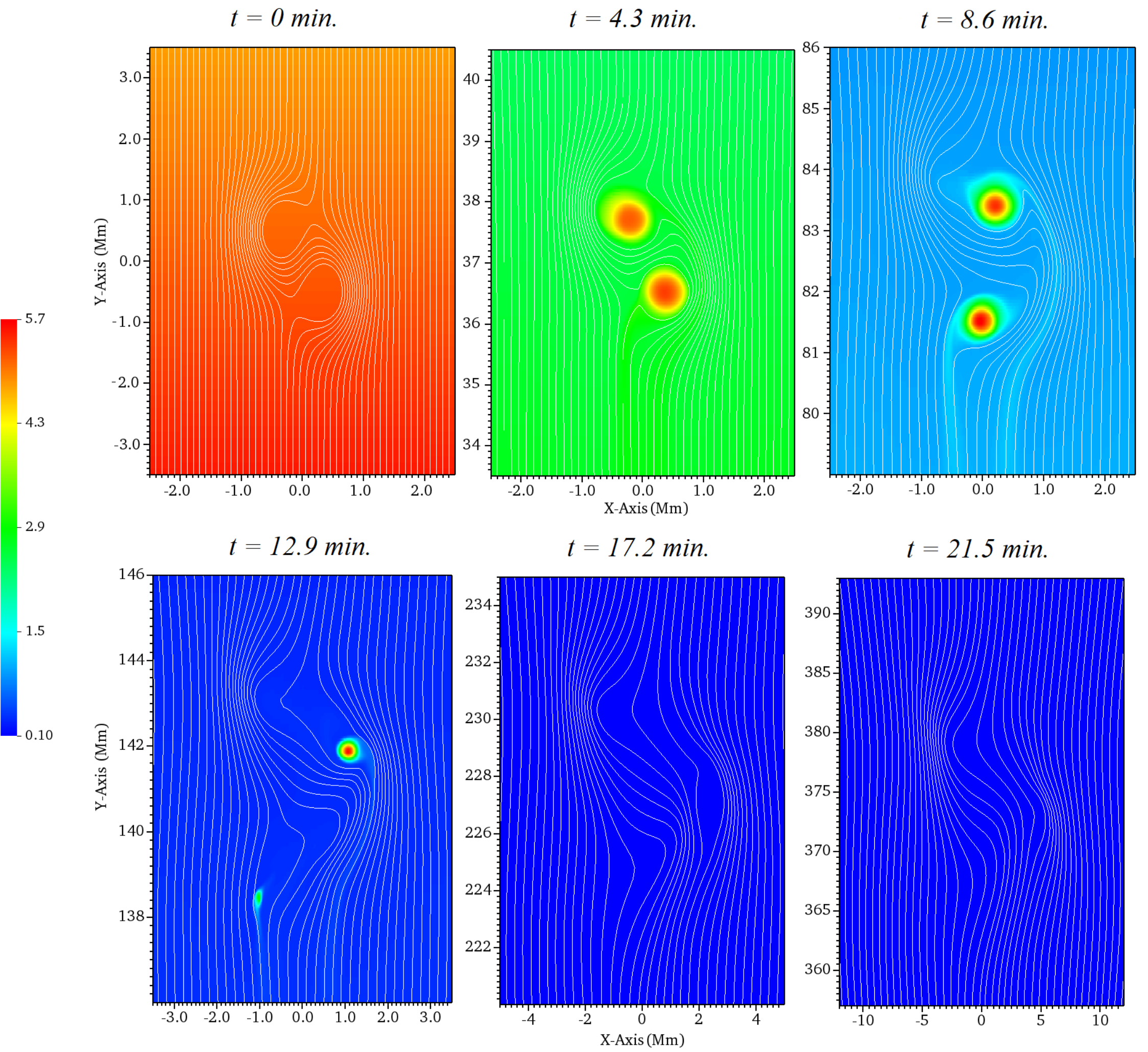} 
        \caption{A sequence of snapshots, showing the evolution of the switchback as it propagates, from left to right, is shown. The magnetic field is critically-expanding ($l_B = 20\ \mathrm{Mm}$). The color plots display density, in units of $2.34 \cdot 10^{-12}\ \mathrm{kg\ m^{-3}}$. Over-plotted are magnetic field lines (a zoom in for a clearer visualization). The leftmost plot depicts the evolution at $t \approx 257\ \mathrm{s}$, and following plots are at equal steps of $\approx 257\ \mathrm{s}$. }
        \label{fig3}
\end{figure}

 
By considering the setup with a critical expansion, we can explore the additional, finite-amplitude effects that contribute to the distortion of the propoagating switchback. The evolution of the switchback propagating through a stratified corona with critical expansion of the magnetic field is presented in Fig.~\ref{fig3}. 
It appears that, despite the lack of WKB-like amplitude changes during propagation in the critically expanding setup, the switchback undergoes deformations, and tends to evolve towards weaker magnetic field deflections. It is clear that magnetic field expansions for which the relative perturbation amplitude of the switchback does not decrease with height contribute to its stability. Nevertheless, the switchback undergoes deformations significantly faster than in a homogeneous plasma \citep{2020ApJS..246...32T}, due to the finite amplitude effects mentioned earlier.
One of these effects is present due to density stratification causing symmetry breaking. To understand this effect, let us investigate the force balance within the switchback solution. The momentum equations reads:
\begin{equation}
\rho \frac{\partial \mathbf{v}}{\partial t} + \rho \left(\mathbf{v} \cdot \nabla\right) \mathbf{v} = -\nabla p + \frac{1}{\mu} \left( \nabla \times \mathbf{B}\right) \times \mathbf{B} - \rho \mathbf{g}.
\label{momeq}
\end{equation}
Let us consider a frame of reference in which the switchback propagating at the Alfv\'en speed is at rest, or in other words, consider a constant background flow\footnote{Strictly speaking, in setups with a varying Alfv\'en speed such an inertial frame of reference does not exist globally, however one could still consider a time-varying background flow in which the switchback would be at rest locally. For now, let us continue with this demonstrative analysis by considering a very slowly varying density for which the constant flow is a good approximation.} $v_{0y} = -V_A = -B_{0y}/\left(\mu \rho \right)^{1/2}$ in the vertical direction. After adding such a background flow, now the total (background and perturbations) velocity and magnetic field obey an Alfv\'enic correlation as in Eq.~\ref{correl}, and not just the perturbations. Then we can replace the expression for the advective derivative in Eq.~\ref{momeq} using Eq.~\ref{correl} for the total velocity field, and we also expand the Lorentz force: 
\begin{equation}
    \rho \frac{\partial \mathbf{v}}{\partial t} + \rho \left(\frac{\mathbf{B}}{\sqrt{\mu\rho}} \cdot \nabla\right) \frac{\mathbf{B}}{\sqrt{\mu\rho}} = -\nabla p + \frac{1}{\mu} \left(\mathbf{B} \cdot \nabla\right) \mathbf{B} - \nabla \frac{B^2}{2\mu} - \rho \mathbf{g}.
    \label{momeq2}
\end{equation}
If the density is constant, that is, if there is no gravitational stratification (implying also that the first and last terms in the right-hand-side of Eq.~\ref{momeq2} vanish), the first term of the Lorentz force cancels exactly with the advective derivative. If the Alfv\'enic wave-package is of constant total magnetic field, i.e. circularly or spherically-polarized, as in the present case, the second term of the Lorentz force also vanishes, and we obtain a stationary flow. This demonstrates that spherically-polarized switchbacks in a uniform medium are a nonlinear wave solution, and explains their relatively stable propagation out to large distances, as in \citet{2020ApJS..246...32T}. Switchbacks propagating in a uniform medium are still susceptible to parametric decay, an instability arising due to random fluctuations of the uniform background \citep{1963SPhD....7..988G,1978ApJ...219..700G}, which is ultimately responsible for their unfolding. 
However, when large-scale density variations are present, as in our case of a gravitaitonally  stratified corona, the terms mentioned previously do not cancel exactly. This is easy to see from the advective derivative in Eq.~\ref{momeq2}, as density cannot be simplified anymore. The resulting inexact cancellation of forces leads to additional deformation of the switchbacks, as these cease to be exact nonlinear solutions. \par
\begin{figure}[h]
    \centering     
        \includegraphics[width=0.65\textwidth]{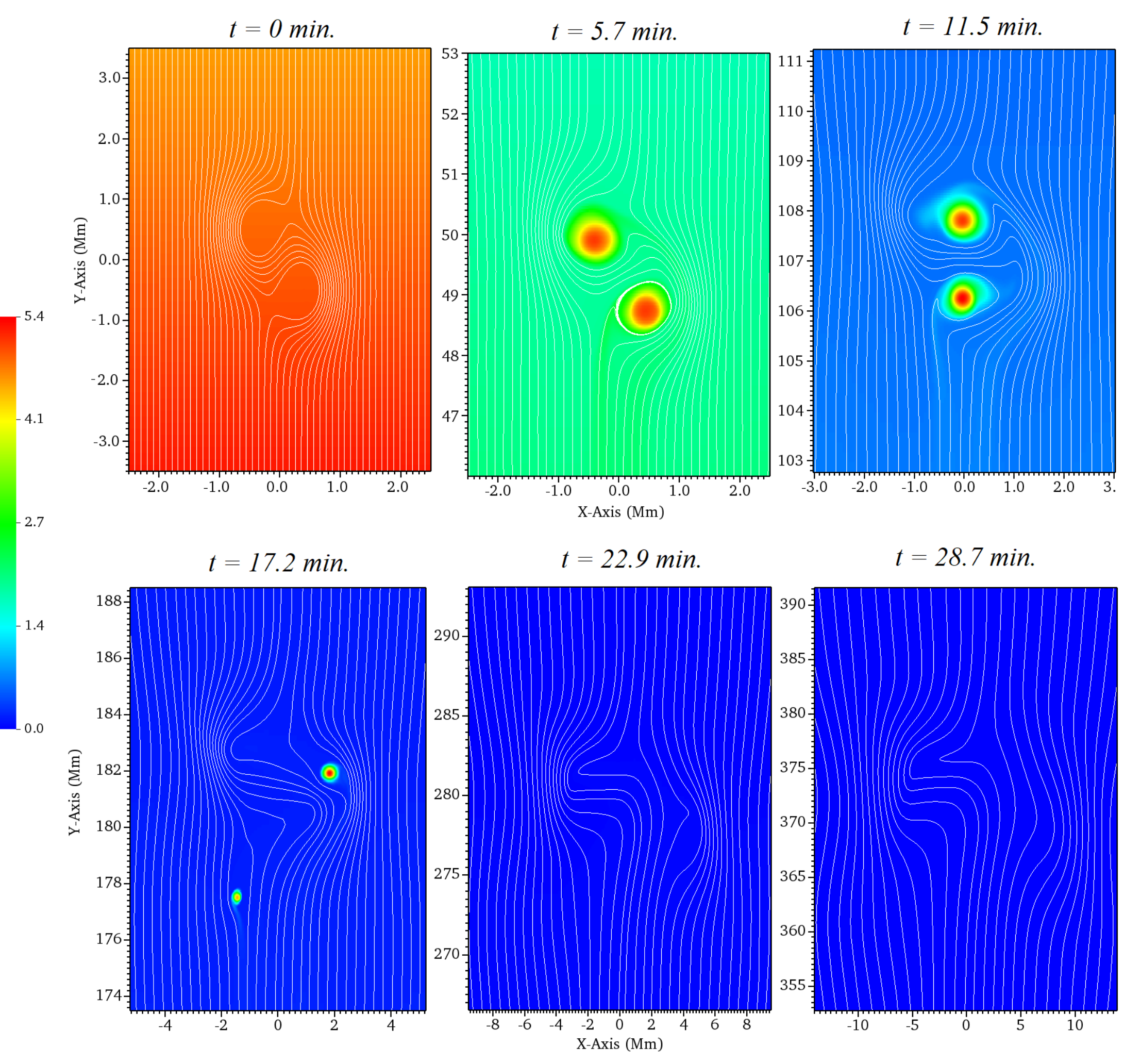} 
        \caption{A sequence of snapshots, showing the evolution of the switchback as it propagates, from left to right, is shown. The magnetic field is expanding super-radially ($l_B = 15\ \mathrm{Mm}$). The color plots display density, in units of $2.34 \cdot 10^{-12}\ \mathrm{kg\ m^{-3}}$. Over-plotted are magnetic field lines (a zoom in for a clearer visualization). The leftmost plot depicts the  evolution at $t \approx 344\ \mathrm{s}$, and following plots are at equal steps of $\approx 344\ \mathrm{s}$. }
        \label{fig3b}
\end{figure}

Finally, let us consider the setup displaying super-radial expansion of the magnetic field, for which the evolution is shown in Fig.~\ref{fig3b}.
It can be observed that while the switchback still undergoes morphological changes, the amount of magnetic field deflection is still significant by the time the switchback propagates out to $400\ \mathrm{Mm}$, i.e. the top radial boundary. While the changes in shape can be attributed to the symmetry-breaking described above, the maintaining of strong deflection comes from the fact that in a super-radially expanding field the relative amplitude of the switchback perturbation grows with radius, as shown in Fig.~\ref{fig1b}. \par 
Besides the effects and forces acting towards unfolding the switchback, there are other important effects present. In all the previous setups (Fig.~\ref{fig2},~\ref{fig3}, and~\ref{fig3b}) we notice that as the switchback propagates upwards, it lifts mass, mostly through the plasmoids which are defined by closed magnetic field lines. The higher density plasma within the closed-field line plasmoids is advected upwards as the plasma is frozen-in. However, as shown in the Appendix, this effect is also there for switchback solutions without plasmoids. For a switchback propagating away from the Sun in the gravitationally stratified solar atmosphere, this leads to the build-up of gravitational potential energy (GPE for short). In the case of e.g., a stone thrown against gravity, a ballistic trajectory results from the stone losing vertical kinetic energy (i.e., slowing down) to GPE.
In the case of a wave package, though, such as the present Alfv\'enic switchback, the gain in GPE cannot come from its gradual deceleration, as it is constrained to propagate at the propagation speed of the medium, in this case the Alfv\'en speed. The excess GPE therefore must come from the magnetic and velocity perturbations of the switchback as the total energy is conserved, that is, resulting in a form of gravitational damping. Although this gravitational damping is not relevant for the coronal setups presented here, as the rise in the GPE is approx. two orders of magnitude smaller than the switchback energy, this is very relevant under chromospheric conditions, as briefly presented in the Appendix.
The mass flow along the propagation direction of the switchback is due to the vertical component of the velocity perturbation, and not due to magnetic pressure gradients, as in the case of ponderomotive forces \citep{1971JGR....76.5155H,1974PhFl...17.1399C}, the magnetic field magnitude being constant. Nevertheless, the discussion above on gravitational damping should apply also for non-circularly polarized waves which lift mass ponderomotively \citep[e.g.,][]{2020ApJ...905..168O}. While ponderomotive forces are present irrespective of gravitational stratification, in the case of the propagating Alfv\'enic switchback there are essentially no density perturbations without gravity. The density perturbation, and thus also the mass uplift, is the result of advection by the velocity perturbation of the switchback in the gravitationally stratified plasma. As the velocity perturbation of the Alfv\'enic switchback is solenoidal, the advection takes place in an essentially incompressible manner, explaining the lack of density perturbations in the constant density case.  Linear Alfv\'en waves do not present perturbations along the propagation direciton, showing the finite-amplitude nature of the effect. \par
\begin{figure}[h]
    \centering     
     \begin{tabular}{@{}cc@{}}
        \includegraphics[width=0.33\textwidth]{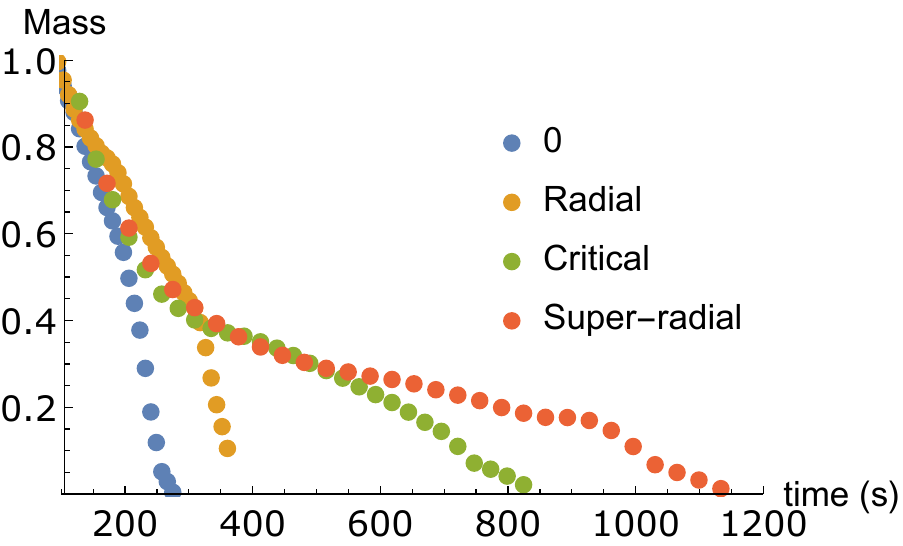}  
        \includegraphics[width=0.33\textwidth]{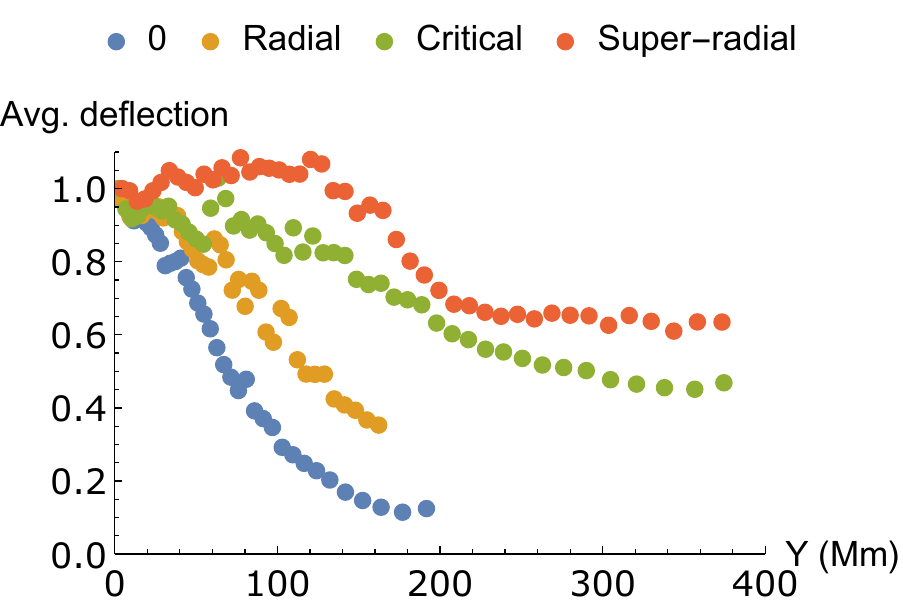}
        \includegraphics[width=0.33\textwidth]{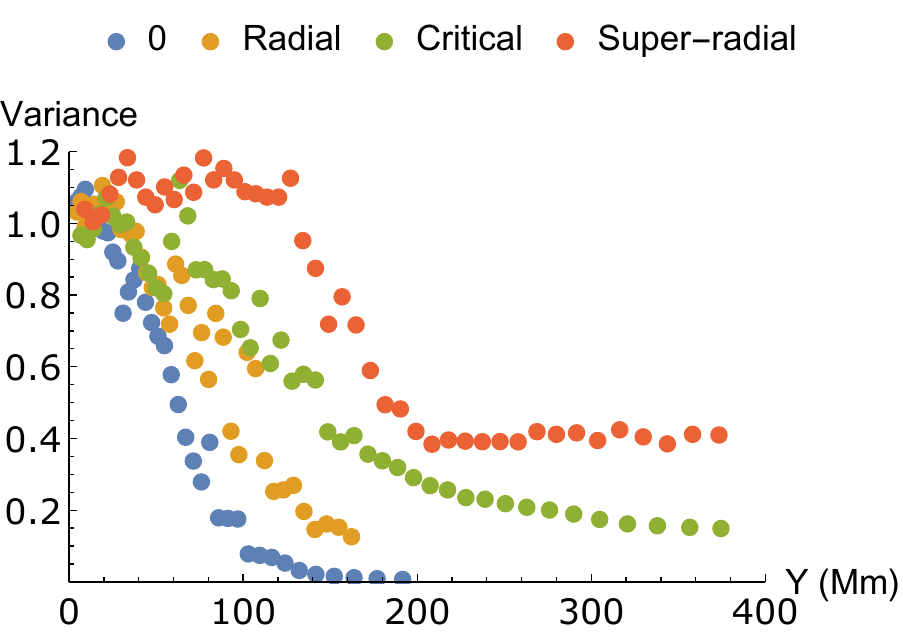} \\
       \end{tabular}    
        \caption{\textit{Left:} Evolution of the mass enclosed in the plasmoids of the propagating switchback. \textit{Center and Right:} The evolution of average deflection angle and variance of the deflection angle, respectively. The data points are at equal time steps which varies between different setups, and the distance at which they are plotted corresponds to the average distance the switchback propagated to. All values are normalized to unity initially.}
        \label{fig4}
\end{figure}

\begin{figure}[h]
    \centering     
        \includegraphics[width=1.0\textwidth]{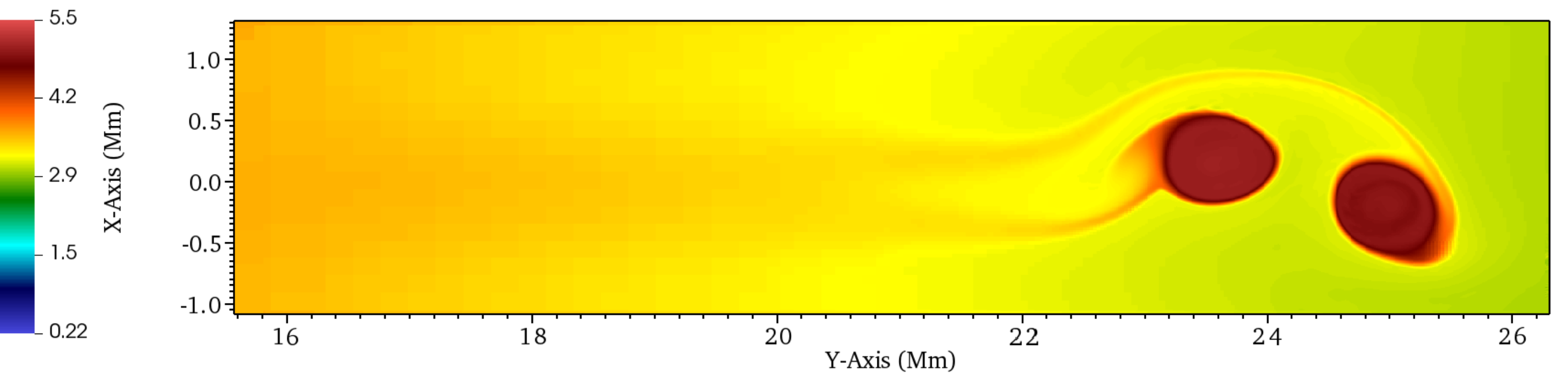} 
        \caption{Snapshot showing the density around the propagating switchback at $t \approx 86\ \mathrm{s}$ for the non-expanding setup. Density is in units of $2.34 \cdot 10^{-12}\ \mathrm{kg\ m^{-3}}$.}
        \label{fig5}
\end{figure}

 
In order to quantify the evolution and deformation of the propagating switchback over time, and compare the setups with different degrees of expansion, in Fig.~\ref{fig4} we show two key statistics on the deflection angle of magnetic field lines passing through and in the vicinity of the switchback, namely the average deflection and the variance of the deflection. The field lines are calculated through \texttt{VisIt}'s \citep{HPV:VisIt} built-in \texttt{IntegralCurve} operator, along which the deflection angles are extracted at discrete intervals. The deflection angle is calculated in the following way: 
\begin{equation}
\alpha = \arctan(|B_x|/B_y)+\frac{\pi}{2}(1-B_y/|B_y|),
\label{defl}
\end{equation}
The average deflection is calculated along each magnetic field line, and then averaged between different field lines:
\begin{equation}
\langle\alpha\rangle = \frac{1}{{\mathit{N}}} \sum_{i}^{N} \alpha_i,\; \alpha_i = \frac{1}{{\mathit{M_i}}}\sum_{j}^{M_i} \alpha_{i,j},
\label{avg}
\end{equation}
where $i,j$ are the indices of field lines and of the different deflection angle values extracted along the $i$-th field line, respectively. $N=40$ is the number of different field lines, integrated from the bottom vertical boundary from equally distanced seeds from $x = -0.5$ to $0.5\ \mathrm{Mm}$, and $M_i$ is the number of extraction points along each field line, determined by the \texttt{IntegralCurve} routine. Extraction points with deflections below 5 degrees are discarded, therefore the different expansions and length of individual field lines do not affect the analysis.
The variance of the average deflection of the field lines is calculated in the following way:
\begin{equation}
\sigma^2 = \frac{1}{N} \sum_i^N \left(\alpha_i - \langle\alpha\rangle  \right)^2
\label{variance}
\end{equation}
The variance indicates the similarity of the deflection between different field lines, and its evolution reflects the deformation of the switchback over time. A variance of zero implies that all magnetic field lines have the same shape. As the individual magnetic field lines originate from outside the switchbacks, these do not include the closed-field-line plasmoids, therefore these do not contribute to this analysis.
Investigating Fig.~\ref{fig4}, we note that the evolution of both the mean and variance are not linear, e.g. in the super-expanding setup the switchback appears to undergo little evolution after a height of $\approx 200\ \mathrm{Mm}$.
There are other effects impacting the evolution of the switchbacks worth mentioning here. At the edge of the plasmoids, parts of the embedded switchback solution, reconnection is continuously operating. The dense plasma appears to drain continuously, as the closed magnetic field lines inside the plasmoids are reconnected to the ambient open magnetic field, leaving a trail of denser plasma. An example of this is shown in Fig.~\ref{fig5}.
Eventually, the plasmoids drain out completely, as even smaller plasmoids are formed in their reconnection process (visible around $y \approx 109\ \mathrm{Mm}$ in Fig.~\ref{fig2}). In the simulations, the reconnection occurs because of numerical resistivity which depends directly on the numerical resolution. We have conducted  a series of refinement studies on the draining rate, i.e., reconnection rate of the plasmoids, and found that the draining rate is saturated after 4 levels of refinement. The reconnection rate is expected to saturate for Lundquist numbers on the order $10^4$ \citep{2012PhPl...19d2303L}, which is the estimated value in our simulations for 4 levels of refinement. Interestingly, the reconnection rate appears to depend on the expansion rate of the magnetic field, the plasmoids surviving further out with increased expansion, as depicted in Fig.~\ref{fig4}. The heights by which the plasmoids drain completely are $\approx 140, 170, 210, 250\ \mathrm{Mm}$, respectively, ordered by increasing expansion. The mass is calculated by multiplying the density inside the plasmoids with their area. The plasmoids are defined by the closed magnetic fields they contain. Note that the evolution of the mass in Fig.~\ref{fig4} is shown a function of time, which is a more accurate way of comparing the different expansion setups, as the Alfv\'en speed varies considerably between these. The draining rate is not constant, and it seems to exhibit breaking points, e.g. around $300\ \mathrm{s}$, where the draining rate accelerates/decelerates for weak/strong expansions, respectively. Furthermore, in simulations without gravity with a similar embedded switchback, although reconnection takes place, it is proceeding at a considerably slower rate than in the simulation with gravity. It is not clear whether this acceleration in reconnection rate is caused by the growing density ratio between the plasmoids and the ambient plasma, or some other effect, but it warrants further investigation. \par 
An additional question regarding the evolution of the switchbacks that should be discussed here is whether the presented results depend on the size of the considered switchback. In this section we have assumed that the WKB approximation is valid for the simulated switchback. This is confirmed by its measured high Alfv\'enicity in the simulation. The WKB approximation holds until the switchback is much smaller than the relevant scale height, in this case the gravitational scale height of $\approx 50 \mathrm{Mm}$. This is expected to be true for supposed switchbacks originating in the solar corona, as otherwise these would be large enough to be routinely observed. However, once the WKB approximation holds, the WKB effects are scale independent, meaning that switchbacks of different sizes are expected to be affected proportionally to the same extent. This is also the case for the finite-amplitude effects, which can be seen in Eq.~\ref{momeq2}: by expanding the advective derivative, the rate of change of the perturbed velocity field can be shown to depend only on the local density gradient. This was verified by carrying out numerical simulations with a 5 times smaller switchback, for the non-expanding and the critically expanding cases, for which a similar evolution was observed. Thus, the presented results hold for all switchbacks smaller than the gravitational scale height in the corona.

\section{Conclusions} \label{sec:concl}

We have conducted a series of simulations of analytical switchback solutions embedded in a 2.5D gravitationally stratified corona and a background magnetic field with varying degrees of expansion. The switchback represents a spherically-polarized Alfv\'enic wave package of constant magnetic field magnitude propagating upwards (i.e., away from the Sun). The evolution of the switchback is followed as it propagates up to 400 Mm. We find that, contrary to a relatively stable propagation in setups without gravity, the switchbacks suffer deformations on faster timescales in the gravitationally stratified corona. With increasing magnetic field expansion rate, the switchback retains stronger deflection. In the case of no magnetic field expansion or radial expansion, the switchbacks unfold into Alfv\'en waves presenting minimal magnetic field deflections by the time they propagate out to the upper radial boundary. The deformation of the switchbacks is determined by multiple, both WKB and finite-amplitude effects, present as a result of gravitational stratification. The WKB approximation holds to a good approximation as the simulated switchback is much smaller than the gravitational scale height. Among the WKB effects, we note wave package stretching along the propagation direction, as across all setups the Alfv\'en speed is increasing with height. Another WKB effect, that of wave energy flux conservation, renders the relative perturbation amplitude of the switchbacks height-dependent, either increasing or decreasing for expansion rates lower or higher than a critical expansion, respectively. In the setup with a critical expansion rate, unaffected by this effect, the switchbacks still evolve into waves showing smaller magnetic field deflections, albeit the remnant deflections are significant compared to the ones in the, e.g. radial expansion setup. Besides the WKB effects, symmetry breaking induced by the gravitational stratification between nonlinear advection and the Lorentz force, which otherwise cancel out exactly in a homogeneous plasma, contributes to the deformation of the initial switchback configuration. When considering super-radial expansion, a property of magnetic fields observed in coronal holes \citep{1976SoPh...49...43K}, the growing relative perturbation amplitude in this setup appears to counteract the unfolding of the switchback through symmetry-breaking, and while pronounced deformations do occur, maximal magnetic field deflections are approximately maintained during propagation. Mass flow in the propagation direction of the switchbacks, across all setups, leads to gravitational potential energy build-up in the stratified atmosphere, as heavier plasma is lifted against gravity. The density perturbation is an Eulerian effect resulting from the conservation of mass in the advected plasma of varying density. This flow is not associated with ponderomotive forces, as there are no magnetic pressure gradients, but is part of the switchback solution. As the switchback is a wave-package constrained to propagate at the propagation speed of the medium, the gain in gravitational potential energy cannot come from its deceleration, and leads to a form of gravitational damping. While this damping is not significant in coronal setting, it is important under chromospheric conditions. The plasmoids, which are part of the analytical switchback solution, are continuously reconnecting with the surrounding open magnetic field, leading to the drainage of the heavier plasma within, which forms an extended thin wake behind the switchback. Refinement studies have shown that the reconnection rate does not depend on the numerical resistivity for Lundquist numbers above $\approx 10^4$, confirming previous studies. However, reconnection appears to proceed faster for weaker magnetic field expansion and at a much faster rate than in a homogeneous plasma without gravity, for reasons which are as of yet not clear to us. \par 
In summary, switchbacks of constant magnetic field magnitude, propagating in a gravitationally stratified plasma with an expanding magnetic field deform at a faster rate than it would result under homogeneous conditions. While for weak magnetic field expansions, up to radial and beyond, the switchbacks unfold into Alfv\'en waves presenting minimal magnetic field deflections, for super-radial expansions observed in coronal holes they retain strong deflections, albeit still showing significant deformations compared to their initial state. In this sense, the degree of magnetic field expansion could act as a filtering criteria for the survivability of switchbacks originating in the lower solar corona out to higher radial distances. In the previous section we noted that in models of the solar corona and solar wind \citep[e.g.,][]{2009ApJ...707.1659C} the maximal Alfv\'en speed is attained around $1.5 - 2\ \mathrm{R}_\odot$, after which the relative magnetic perturbation amplitude is increasing with radial distance, even for radial expansions. In this sense, even the very weak magnetic field deflections to which switchbacks unfold in the radially expanding magnetic field setup will tend to evolve into high-amplitude Alfv\'en waves presenting strong deflections. However, the  formation of switchbacks in this way would constitute an in-situ mechanism \citep{2020ApJ...891L...2S,2021arXiv210109529S}, and would not require the generation of switchbacks in the lower solar atmosphere, but only of small amplitude Alfv\'en waves. Therefore, for switchbacks to be considered originating in the lower solar atmosphere it is implied that these structures formed in the corona propagate out into the solar wind largely unaltered, or that they display some properties which cannot be explained by in-situ generation mechanisms, but by e.g., interchange reconnection. Based on the presented simulations, it is unlikely that switchbacks originating in the lower solar corona propagate out into the solar wind largely unaltered. Thus, if switchbacks are indeed propagating Alfv\'enic wave-packages, their remnants which escape into the solar wind would probably be indistinguishable from in-situ steepened Alfv\'en waves, unless their generation mechanism, e.g. through reconnection, imprints the switchbacks with distinct properties not seen in Alfv\'en waves. \par 
The present study has a number of caveats, however. Since to the best of our knowledge there are no known analytical and localized 3D solutions of switchbacks of constant magnetic field magnitude, we had to limit our study to a 2.5D simulation. Numerical solutions in 3D do exist, however, these are for periodic domains \citep{2019ApJ...881L...5V}. It is unclear to what degree the evolution of a three-dimensional switchback would differ from the one presented here. Perhaps one advantage of a 3D study would be a differentiation of switchbacks depending on the nature of the plasmoids they contain. In the present study, these could represent the cross-sections of flux ropes connecting to the lower atmosphere, torus-like closed flux ropes, or globular plasmoids. The evolution of the switchback, including the reconnection rate it undergoes is probably different for these different cases. Moreover, the reconnection rate might be slowed down by effects not considered in the present study, such as velocity shears \citep{https://doi.org/10.1029/96JA03144}, or diamagnetic stabilization \citep{2010ApJ...710.1769S,2010ApJ...719L.199P,2013GeoRL..40...11P}. If the reconnection rate is indeed slower than it appears here, the plasmoids could propagate out to longer distances before completely draining out. However, this scenario raises another question: In cases when the switchbacks unfolds, such as for radial expansion, the plasmoids appear to detach from the Alfv\'en wave envelope that was advecting them, and start following ballistic trajectories. This is not observed in our simulations, as plasmoids drain before they detach, however, it is shown to be the case for chromospheric propagation, in the Appendix. Another shortcoming of the present study is the artificial embedding of the switchback solution on a background magnetic field. A self-consistent generation of switchbacks from interchange reconnection in the corona within a 3D numerical study, including following the evolution of the resulting switchbacks, is a possible future improvement to the current study. Although we did not consider non-ideal energy sources (thermal conduction, radiative losses, etc.), the low plasma-beta corona is mostly governed by magnetic forces. Nevertheless, non-ideal terms could impact the evolution of the uplifted plasma inside the plasmoids, and their reconnection rate.

\appendix

In order to keep the main body of this paper more transparent, simulations with different switchback parameters and conditions are presented in this Appendix. These additional simulations are carried out in a non-expanding uniform magnetic field. The results strengthen our conclusions on the property of switchbacks propagating through a gravitationally stratified corona, namely the significant deformations these undergo, and still display the other effects described in Section~\ref{sec:result}.

\section{Switchback propagation without gravity}

\begin{figure}[h]
    \centering     
        \includegraphics[width=0.5\textwidth]{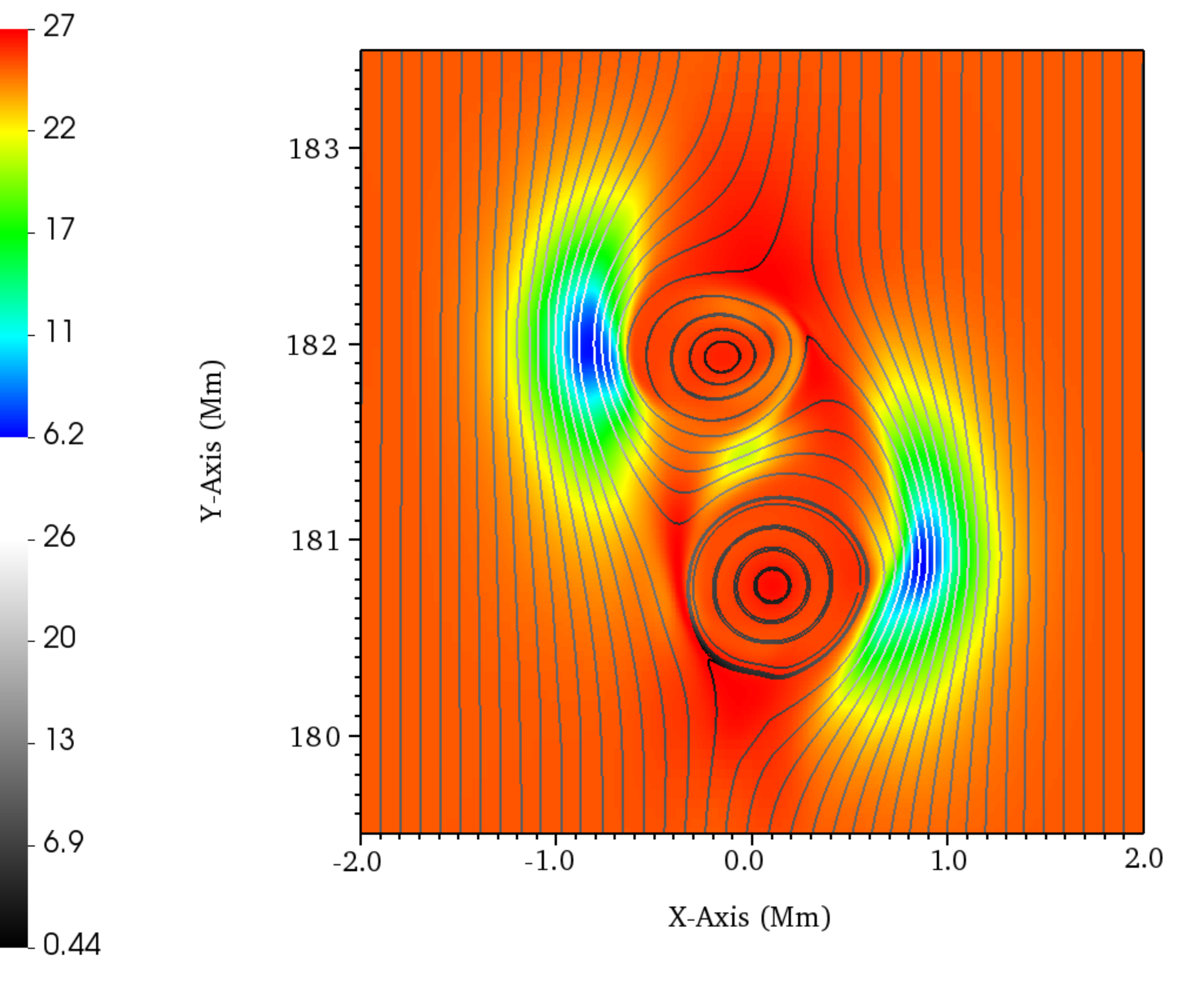}
        \caption{Snapshot depicting the magnetic field configuration of the switchback, as it appears near the top boundary. Field lines of the in-plane ($x-z$-plane) magnetic field are shown, with the gray-scale bar representing the in-plane magnetic field magnitude. The color plot shows the out-of-plane component of the magnetic field. Magnetic field is in units of Gauss.}
        \label{App1}
\end{figure}

In this simulation, the setup differs from the ones presented in Section~\ref{sec:model} in setting a constant density $(\rho = 4.68 \cdot 10^{-12}\ \mathrm{kg\ m^{-3}})$ and pressure (for $T = 1\ \mathrm{MK}$) throughout the domain, and turning gravity off. In this sense, our setup is similar to the one in \citet{2020ApJS..246...32T}, except for the non-periodic domain. While the initial shape of the switchback is the same as in Fig.~\ref{fig1}, see in Fig.~\ref{App1} as it appears at the end of the simulation.
Our results qualitatively agree with the findings of \citet{2020ApJS..246...32T}, in that the switchback appears to be relatively stable as it propagates, and minimal deviations occur over the full length of the simulation domain. These deviations, and ultimately their unfolding are the result of the parametric decay instability, as mentioned previously.

\section{Switchback with symmetrically-positioned plasmoids}

\begin{figure}[h]
    \centering     
        \includegraphics[width=1.0\textwidth]{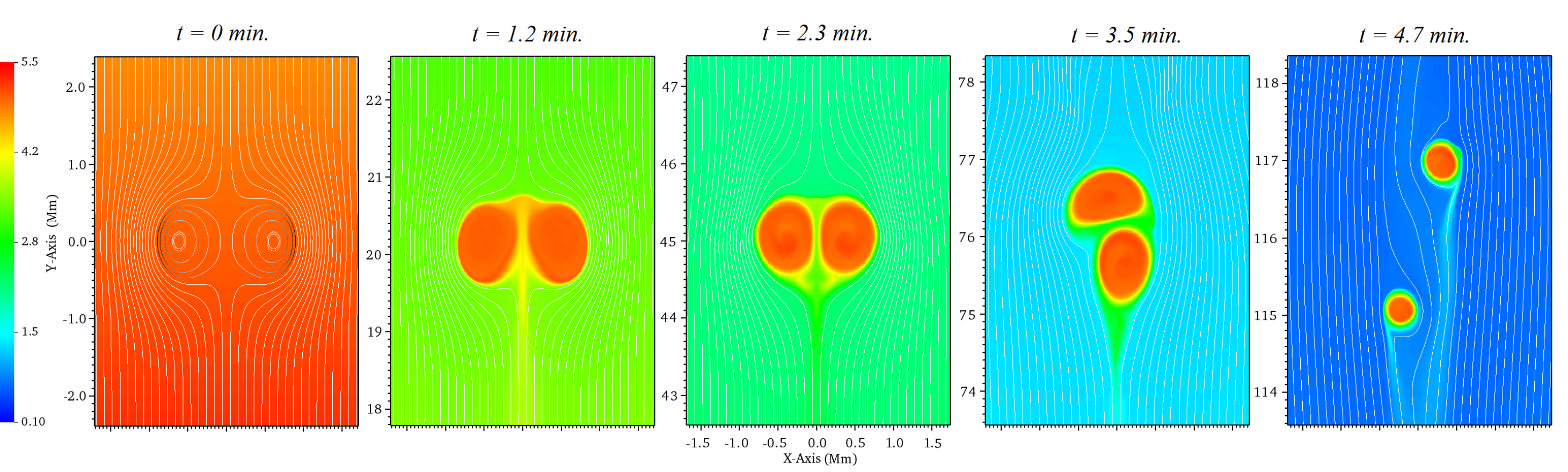}
        \caption{A sequence of snapshots following the evolution of the switchback as it propagates, from left to right. The color plots show density, in units of $2.34 \cdot 10^{-12}\ \mathrm{kg\ m^{-3}}$. Over-plotted are magnetic field lines (zoom in for a clearer visualization). The leftmost plot depicts the initial condition, and following plots are at equal steps of $\approx 86\ \mathrm{s}$.}
        \label{App2}
\end{figure}

In Section~\ref{sec:model} it is argued that the switchback wave package undergoes vertical stretching as it propagates, given there is an Alfv\'en speed gradient in the vertical direction. Our choice for the asymmetrical shape of the switchback was motivated by a better comparison to the existing simulations. However, the question remains whether a switchback with symmetrically-positioned plasmoids could remain stable for a longer duration. In this sense, we set now $y_1 = 0$ and $y_2 = 0$ in Eqs.~\ref{bpert}, in the otherwise unchanged model described in Section~\ref{sec:model}, while still without magnetic field expansion. The evolution of the symmetrical switchback as it propagates is presented in Fig.~\ref{App2}.
The symmetrically positioned plasmoids are an unstable configuration, as it appears, and eventually flip over one another as the switchback propagates, after which the dynamics are similar to the one presented in Fig.~\ref{fig2}: the plasmoids gradually drain out through reconnection and the switchback unfolds.

\section{Switchback without plasmoids}

\begin{figure}[h]
    \centering     
        \includegraphics[width=1.0\textwidth]{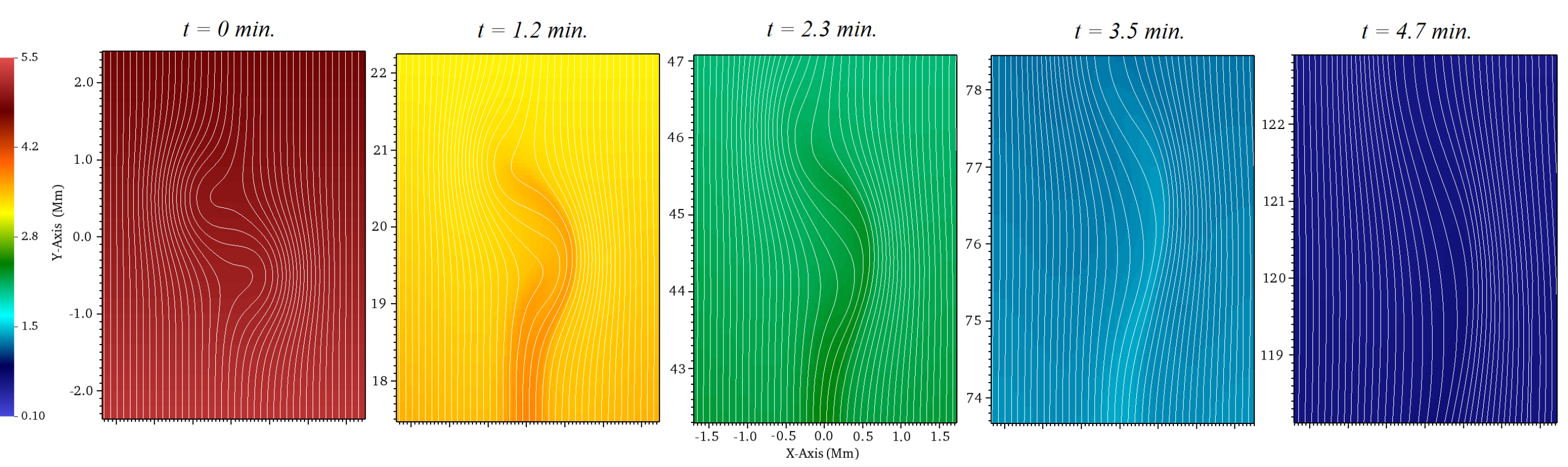}
        \caption{Same as in Fig.~\ref{App2}, but for the switchback without plasmoids.}
        \label{App3}
\end{figure}

In the Introduction we have described that plasmoids appear to be integral parts of the switchback phenomenon with deflections strong enough to cause polarity reversals, both in the present analytical solution and in numerical results. However, switchback solutions being without plasmoids present might still exist. Therefore, it is essential to establish whether the effects discussed in Section~\ref{sec:result} still apply to switchbacks which do not present embedded plasmoids, such as the observed mass transport. 
Switchback solutions without embedded plasmoids can readily be obtained from Eq.~\ref{bpert}, by reducing the amplitude of the perturbation, setting $A = 5\ \mathrm{G \cdot Mm}$. Otherwise the setup is similar to the one described in Section~\ref{sec:model}. The results of the simulation are presented in Fig.~\ref{App3}. 
In the evolution of the plasmoid-less switchback, one can identify the main effects and mechanisms as are described in Section~\ref{sec:result}. Mass transport is still present, although to a lesser extent. Since there are no closed field lines, mass flows are allowed and therefore the drainage of the uplifted mass is more efficient. On the other hand, the stretching of the switchback and the reduction in the amplitude of the magnetic field perturbation, in accordance with the conservation of wave energy flux, are qualitatively the same. Note that for super-radial expansions, for which the relative perturbation amplitude of the switchback grows with radial distance, the reverse would be observed, leading to stronger deflections. 

\section{Switchback embedded in the chromosphere}

\begin{figure}[h]
    \centering     
        \includegraphics[width=1.0\textwidth]{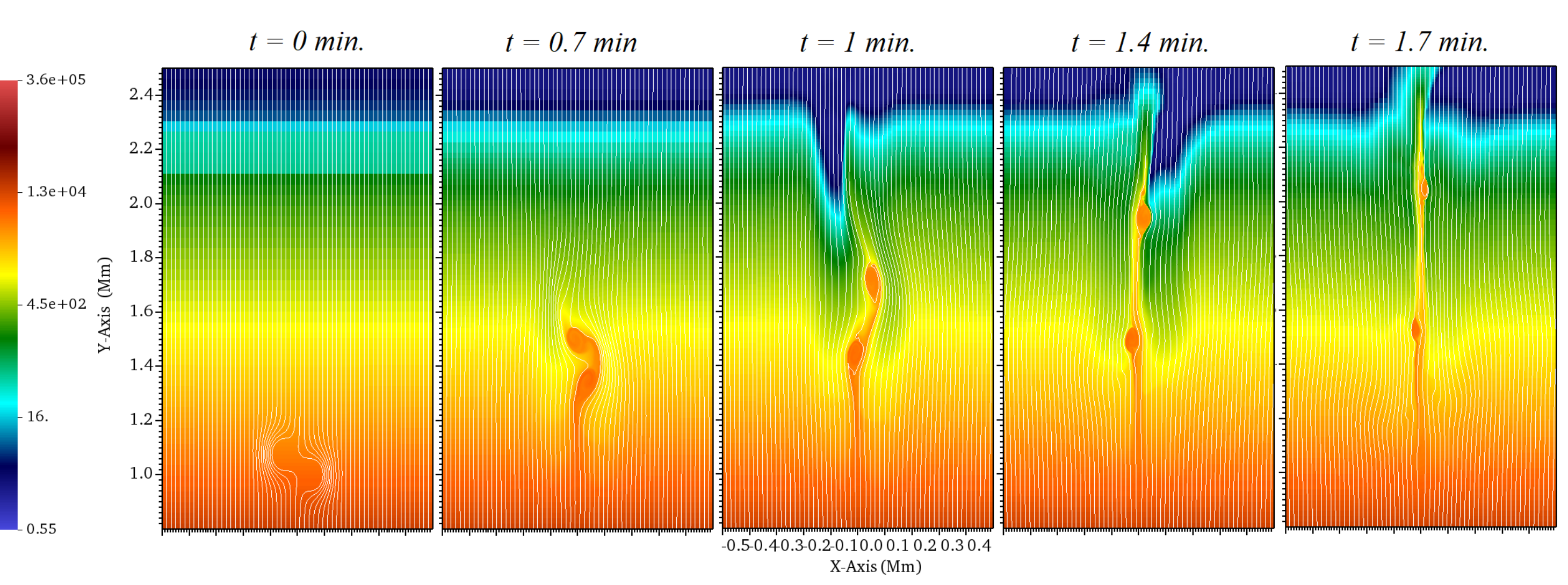}
        \caption{A sequence of snapshots presenting the evolution of the switchback, from left to right. The color plot show density, in units of $2.34 \cdot 10^{-12}\ \mathrm{kg\ m^{-3}}$, in logarithmic scale. Overplotted are magnetic field lines (zoom in for a clearer visualization). The time stamp of each snapshot is shown above.}
        \label{App4}
\end{figure}

In Paper I, we have investigated whether transverse or up-flows, in and around e.g. at footpoints of photospheric magnetic bright points can lead to switchbacks, and whether these are able to propagate out freely into the corona. While the answer to the latter question is negative, it is still of interest in this context to investigate the evolution of a switchback embedded in the chromosphere. Therefore we modified the setup described in Section~\ref{sec:model} in order to include a more realistic lower solar atmosphere, using the VAL-C temperature profile \citep{1981ApJS...45..635V}. The shape of the embedded switchback is similar to the one described in Section~\ref{sec:model}, however, its vertical extent is reduced to $\approx 200\ \mathrm{km}$, embedded at a height of $1\ \mathrm{Mm}$ above the photosphere. In the current model, at this height, the plasma-beta is already below unity. Otherwise the setup is similar, with no expansion of the magnetic field. The evolution of the switchback is depicted in Fig.~\ref{App4}.
Initially, the dynamics are comparable to the one presented in Fig.~\ref{fig2}, such that the switchback is undergoing stretching and unfolding, the switchback is lifting up mass, and mass is draining from the embedded plasmoids through reconnection. However, these processes take place much faster, that is, over a much shorter distance than in the corona. This is due to the much shorter scale height of around $\approx 300\ \mathrm{km}$ at the initial position of the switchback. At later times, as the leading edge of the switchback reaches transition region heights (at around $2.3\ \mathrm{km}$), it perturbs the layer, resulting in dips and spikes. Eventually the magnetic field straightens out, with the remaining plasmoids draining. However, highly folded magnetic field lines do not enter the corona, in agreement with our conclusions in Paper I. Note that in Paper I magnetic field expansion was self-consistently accounted for, unlike in the present case. An important difference here compared to the coronal propagation is that the gravitational potential energy density is much higher, and it equals the wave energy density over an uplift of $\approx 400\ \mathrm{km}$. Therefore the gravitational damping of the switchback is significant, and it has a strong impact on the unfolding of the switchback. It appears that once the switchback unfolds, resulting in a largely straight magnetic field, the plasmoids detach from the wave package and decelerate, as the wave continues propagating upwards. We have repeated the chromospheric simulation with a plasmoid-less switchback as well, with relatively similar results as for the coronal counterpart.

\acknowledgments

N.M. was supported by a Newton International Fellowship of the Royal Society.
D.U. is thankful for the support received through FWF project P27800. This research has received financial support from the European Union’s Horizon 2020 research and innovation program under grant agreement No. 824135 (SOLARNET) enabling D.U. a visit to Sheffield University.
R.E. is grateful to Science and Technology Facilities Council (STFC, grant number ST/M000826/1) UK and the Royal
Society for enabling this research. R.E. also acknowledges the support received by the CAS Presidents International
Fellowship Initiative Grant No. 2019VMA052 and the warm hospitality received at USTC of CAS, Hefei, where part of his contribution was made.
V.M.N. acknowledges the STFC consolidated grant ST/T000252/1 , and the BK21 plus program through the National Research Foundation funded by the Ministry of Education of Korea.

\bibliography{Biblio}{}
\bibliographystyle{aasjournal}

\end{document}